\documentclass[prl,aps,preprint]{revtex4}

\def\be{\begin{equation}}
\def\ee{\end{equation}}
\def\bear{\begin{eqnarray}}
\def\eear{\end{eqnarray}}
\def\beqn{\begin{eqnarray}}
\def\eeqn{\end{eqnarray}}

\begin{document}
%\draft
%\setlength{\baselineskip}{20pt}
%\date{\today}
\preprint{\vbox{\baselineskip=12pt
\rightline{MADPH-08-1525}
}}
\title{Icosahedral ($A_5$) Family Symmetry and the Golden Ratio Prediction for Solar Neutrino Mixing} 
\author{
 Lisa~L.~Everett and Alexander~J.~Stuart}
\affiliation{Department of Physics, University of Wisconsin,
Madison, WI 53706, USA}

\begin{abstract}
We investigate the possibility of using icosahedral symmetry as a family symmetry group in the lepton sector.  The rotational icosahedral group, which is isomorphic to $A_5$, the alternating group of five elements, provides a natural context in which to explore (among other possibilities) the intriguing hypothesis that the solar neutrino mixing angle is governed by the golden ratio, $\phi=(1+\sqrt{5})/2$.  We present a basic toolbox for model-building using icosahedral symmetry, including explicit representation matrices and tensor product rules.  As a simple application, we construct a minimal model at tree level in which the solar angle is related to the golden ratio, the atmospheric angle is maximal, and the reactor angle vanishes to leading order.  The approach provides a rich setting in which to investigate the flavor puzzle of the Standard Model.
\end{abstract}
\maketitle
%\vskip 0.3truecm
%\pacs{\tt PACS number(s): }
%\vskip2cm
%\leftline{} 
%\pacs{}
%\newpage
\section{Introduction}
\label{sI}
The discovery \cite{Fukuda:1998mi} and subsequent measurements \cite{Collaboration:2007zza,sno,
Fukuda:2002pe,Eguchi:2002dm,Apollonio:1999ae,Boehm:1999gk,Ahn:2002up} of neutrino oscillations have verified that neutrinos are massive and lepton mixing is observable, the first particle physics evidence of physics beyond the Standard Model (SM). The data have revealed an intriguing pattern of lepton flavor mixing parameters,  providing a new facet to the SM flavor puzzle.  Current measurements do not allow for an unambiguous extraction of the neutrino mass pattern.  However, global analyses of the data \cite{global} have provided the following $3\sigma$ limits on the mass-squared differences $\Delta m^2_{ij}\equiv m^2_j-m^2_i$ ($m_i$ are the neutrino mass eigenstates):
\begin{equation}
\label{msqsol}
\Delta m^2_{12}=7.67^{+0.67}_{-0.61}\times 10^{-5} \,{\rm eV}^2,
\end{equation}
and
\begin{equation}
\label{msqatm}
\Delta m^2_{13}=\Bigg \{ \begin{array}{c} -2.37^{+0.43}_{-0.46} \times 10^{-3}\,{\rm eV}^2\; {\rm (inverted \,hierarchy)}\\ 2.46^{+0.47}_{-0.42} \times 10^{-3}\,{\rm eV}^2\; {\rm (normal\, hierarchy)}.  \end{array} 
\end{equation}
Therefore, not only is there a suppression of the neutrino mass scale with respect to the electroweak scale, which is known both from direct bounds \cite{pdg} and cosmological constraints \cite{cosmo}, but the neutrino oscillation data also suggest a relatively mild neutrino mass hierarchy as compared to the mass hierarchies of the charged fermions. 
 
The mixing angles of the  Maki-Nakagawa-Sakata-Pontecorvo (MNSP) \cite{Maki:1962mu,Pontecorvo:1967fh} lepton mixing matrix are also constrained to the following $3\sigma$ ranges (in degrees):
\begin{equation}
\theta_{12}=34.5^{+4.8}_{-4.0},\;\;\; \theta_{23}=42.3^{+11.3}_{-7.7},\;\;\; \theta_{13}=0.0^{+12.9}_{-0.0}.
\end{equation}
In contrast to the small angles of the  Cabibbo-Kobayashi-Maskawa (CKM) \cite{ckm} quark mixing matrix, the MNSP matrix has two large angles, the solar angle ($\theta_{12}$) and the atmospheric angle ($\theta_{23}$), and only one small angle, the reactor mixing angle $\theta_{13}$, which is bounded to be less than the quark Cabibbo angle, the largest angle of the quark sector.

To understand the origin of fermion masses and mixings in the SM, the paradigm is to use family symmetries to restrict the form of the mass matrices, and to generate mass hierarchies and mixings through small symmetry breaking perturbations.   For the lepton sector, it is standard to work within the the framework of the seesaw mechanism \cite{seesaw}, which provides arguably the most natural explanation for the suppressed neutrino mass scale.  In terms of mixing angles, it is natural to perturb about vanishing mixings for the quarks.  However, the presence of large angles in the lepton sector hints at a range of possible theoretical starting points. The data suggest that at leading order, $\theta_{13}$ is likely to be zero and $\theta_{23}$ is likely to be maximal ($45^{\circ}$), but there are a range of possibilities  for $\theta_{12}$.  The approach known as ``quark-lepton complementarity'' \cite{qlc}, which has been widely studied in the literature, assumes that $\theta_{12}$ is also maximal at leading order, but is shifted to values within the experimental range by Cabibbo-sized effects.  A perhaps even more popular approach is to seek the Harrison-Perkins-Scott (HPS) ``tri-bimaximal'' mixing pattern \cite{hps} at leading order, in which 
\begin{equation}
\theta_{12}=\theta_{12}^{({\rm HPS})}=\tan^{-1}\left (\frac{1}{\sqrt{2}} \right ) = 35.26^\circ,
\end{equation}
well within the allowed range of the data.  An attractive feature of the tri-bimaximal mixing scenario is that it emerges quite naturally from flavor theories based on non-Abelian discrete family symmetries which are subgroups of $SO(3)$ or $SU(3)$.  Models have been constructed based on $\mathcal{A}_4$ (which is isomorphic to the tetrahedral symmetry group $\mathcal{T}$) \cite{A4},  the binary tetrahedral group ${}^{(2)}\mathcal{T}$ \cite{binarytetrahedral},  $\Delta(3n^2)$ \cite{deltagrp}, the semidirect product of $\mathcal{Z}_3$ and $\mathcal{Z}_7$ \cite{ramond}, $\mathcal{S}_4$ \cite{S4} (which is isomorphic to the  rotational symmetry of the cube),  and $\mathcal{S}_3$ \cite{S3} (here additional input is needed to achieve the HPS value of $\theta_{12}$), among others.  The plethora of viable flavor models within this approach (see \cite{reviews} for reviews) suggests that discrete non-Abelian family symmetries may be the key to understanding the flavor patterns of the lepton sector.

In this paper, we explore the possibility that the rotational icosahedral symmetry group $\mathcal{I}$, which is isomorphic to $\mathcal{A}_5$, the alternating group of five elements, is the family symmetry group in the lepton sector.  Icosahedral symmetry is the only symmetry based on Plato's five ``perfect'' solids (the tetrahedron, cube, octahedron, dodecahedron, and icosahedron) which has not been used as a family symmetry group (the cube and the octahedron are described by the same symmetry group, as are the dodecahedron and icosahedron).   This group has remained comparatively unexplored for physics applications, at least in part because it is not a crystallographic point group.  Icosahedral symmetry is a logical candidate for a family symmetry, in that it has irreducible triplet representations  (as do the groups given above, except for $\mathcal{S}_3$), and it contains the tetrahedral ($\mathcal{A}_4$) group as a subgroup.

An intriguing feature of the icosahedral ($\mathcal{A}_5$) symmetry group is that  it provides a natural setting in which to explore the idea \cite{Datta:2003qg,Kajiyama:2007gx} that the solar mixing angle is related at leading order to the golden ratio, $\phi=(1+\sqrt{5})/2$ (a solution of  $x^2-x-1=0$):
\begin{equation}
\label{goldenprediction}
\theta_{12}=\tan^{-1}\left (\frac{1}{\phi} \right ) = 31.72^\circ,
\end{equation}
which is $2\sigma$ below the current best fit value (but can easily be shifted to higher values by small perturbations).  This hypothesis was first mentioned in \cite{Datta:2003qg}, and recently explored in the context of $\mathcal{Z}_2\times \mathcal{Z}_2$ models in \cite{Kajiyama:2007gx} (it is worth noting that an alternative pattern in which $\cos\theta_{12}=\phi/2$ has also been suggested \cite{rodejohann1}).  It was suggested in  \cite{Kajiyama:2007gx} that $\mathcal{A}_5$ might naturally generate this pattern, as the golden ratio is related to the geometry of the icosahedron. 

In this paper, we investigate the rotational symmetry group of the icosahedron as a family symmetry group, with the specific goal of building flavor models which result in the golden ratio prediction for the solar mixing angle as in Eq.~(\ref{goldenprediction}).   To this end, we  begin by reviewing basic properties of the rotational icosahedral symmetry group.  We will identify a particularly useful group representation for our purposes, first given in the literature by Shirai \cite{shirai}, and use it to construct tensor products of the irreducible representations. Finally, we apply this formalism by constructing toy flavor models which result in Eq.~(\ref{goldenprediction}) at leading order.  Our studies will demonstrate that the icosahedral ($\mathcal{A}_5$) symmetry group provides a rich setting in which to explore the flavor puzzle of the SM.

\section{Theoretical Background}
\label{sII}
In this section, we will provide a brief overview of the basic properties of the icosahedral symmetry group.  The methodology and many of the basic group properties are well known in the literature and can be found in \cite{backhousegard,Coxeter,Hamermesh,Lomont,ramondbk,cumminspatera,shirai,hoyle,luhn}.  However, the explicit tensor product decomposition has not to our knowledge been presented in the literature, at least not for the specific group representation (due to Shirai \cite{shirai}) that we will argue is particularly well suited for the purposes of flavor model building.  We compute these quantities at the end of this section.  \\

%\subsection{Preliminaries}
\noindent {\bf Preliminaries.} The rotational icosahedral group, ${\mathcal{I}}$, is the group of all rotations of the icosahedron which take vertices to vertices. 
%This group, which is  a subgroup of $SO(3)$, is isomorphic to $\mathcal{A}_5$, the group of even permutations of five objects.   Furthermore, icosahedral symmetry is incompatible with translational symmetry, so $\mathcal{I}$ is not a crystallographic point group.  
Recalling that the icosahedron is the Platonic solid which consists of 20 equilateral triangles, such that it has 20 faces, 30 edges, and 12 vertices, it is straightforward to see that there are five types of such rotations (we consider only proper rotations, and therefore ignore inversions).  These rotations include rotations by 0 or $2 \pi$ (the identity), rotations by $2\pi/5$ and $4\pi/5$ about an axis through each vertex, rotations by $2\pi/3$ about axes through the center of each face, and rotations by $\pi$ about  the midpoint of each edge.  These five types of rotations form five disjoint conjugacy  classes, which are denoted as follows: 
\begin{equation}
1, \,12C_5, \,12C^2_5,\, 20 C_3,\, {\rm and} \, 15C_2.
\end{equation}
In the above, we follow the standard practice of using Schoenflies notation, in which 
${C}^{k}_{n}$ is a rotation by $2k\pi/n$, and the number in front identifies the number of group elements in the conjugacy class.  It is worth noting here that this implies that elements of $C^{k}_n$ are order $n$ elements (elements which result in the identity after $n$ operations).  As can be seen from this classification, the rotational icosahedral group has 60 elements (it is order 60).  Given that the order of the group is equal to the sum of the squares of the dimensions of the irreducible representations, and furthermore that the number of conjugacy classes is equal to the number of irreducible representations, the dimensions of the irreducible representations are given by
\begin{equation}
 1+12+12+15+20=60=1^2+3^2+3^2+4^2+5^2.
 \end{equation}
Hence, $\mathcal{I}$ has five irreducible representations: 1, 3, $3^{\prime}$, 4, and 5 (note the presence of two distinct triplet representations, 3 and $3^\prime$).  The single one-dimensional representation, as opposed to the case of $\mathcal{A}_4$, which has three one-dimensional irreducible representations, results because $\mathcal{I}$ is a perfect group (it is equal to its commutator subgroup) \cite{Lomont,ramondbk}. 
 
The character table of the rotational icosahedral group is given in Table I (this table can also be found in \cite{Coxeter,Lomont,backhousegard,cumminspatera,hoyle,shirai,luhn}).  The character table labels the character (trace) of the matrices which represent the group elements of a particular conjugacy class within a given irreducible representation.   Table I indicates the appearance of the golden ratio for the triplet representations of the $12C_5$ and $12C_5^2$ conjugacy classes.  
\begin{table}
\begin{tabular}{|c|c|c|c|c|c|}
\hline
\textbf{$\mathcal{I}$}&\textbf{1}&\textbf{3}&$\textbf{3}^{\prime}$&\textbf{4}&\textbf{5}\\ \hline
\textbf{$e$}&1&3&3&4&5\\ \hline
\textbf{$12 C_5$}&1&$\phi$&$1-\phi$&-1&0\\ \hline
$12{C}^2_{5}$&1&$1-\phi$&$\phi$&-1&0\\ \hline
$20{C}_{3}$&1&0&0&1&-1\\ \hline
$15 {C}_{2}$&1&-1&-1&0&1 \\ \hline
\end{tabular}
\label{table:Char.Tab.}
\caption{The character table of the icosahedral symmetry group $\mathcal{I}$ and the alternating group $\mathcal{A}_5$, in which  $\phi=(1 + \sqrt{5})/2$.}
\end{table}
Using the character table, the decomposition of tensor products of various irreducible representations can be deduced.  More explicitly, the product of two three-dimensional representations $3\otimes 3=1\oplus 3\oplus 5$ (and similarly for  $3^\prime \otimes 3^\prime$, with $3\rightarrow 3^\prime$).  The full list of tensor products is presented in Table II (it also appeared earlier in the literature \cite{luhn}).  Given the properties of tensor products of $SO(3)$, the tensor product decomposition of two elements of the same irreducible representation can be classified by symmetry properties.  For example, $3\otimes 3=1\oplus 3\oplus 5= (1\oplus 5)_S\oplus 3_A$, in which the labels correspond to symmetric or antisymmetric, respectively (see \cite{luhn} for the complete listing). 
\begin{table}
\label{table:Kron. Prod}
\begin{tabular}{|c|}
\hline
$3 \otimes 3$=$1 \oplus 3 \oplus 5$\\ 
$3^{\prime} \otimes 3^{\prime}$ =$ 1 \oplus 3^{\prime} \oplus 5$\\
$3 \otimes 3^{\prime}$=$4 \oplus 5$ \\
$3 \otimes 4$=$3^{\prime} \oplus 4 \oplus 5$ \\
$3^{\prime} \otimes 4$=$3\oplus 4 \oplus 5$ \\
$3 \otimes 5$=$3 \oplus 3^{\prime} \oplus 4 \oplus 5 $\\
$3^{\prime} \otimes 5$=$3 \oplus 3^{\prime} \oplus 4 \oplus 5 $\\
$4 \otimes 4$=$1 \oplus 3 \oplus 3^{\prime} \oplus 4 \oplus 5$\\
$4 \otimes 5$=$3 \oplus 3^{\prime} \oplus 4 \oplus 5 \oplus 5 $\\
$5 \otimes 5$=$1 \oplus 3 \oplus 3^{\prime}\oplus 4 \oplus 4 \oplus 5 \oplus 5$\\
\hline
\end{tabular}
\caption{The tensor product decomposition for the rotational icosahedral symmetry group $\mathcal{I}$  and the alternating group $\mathcal{A}_5$;  see also \cite{luhn}.}
\end{table}
We also show in Table III the relations between the irreducible representations of $\mathcal{I}$ and the irreducible representations of $\mathcal{A}_4$, which are also found in \cite{cumminspatera}.\\
\begin{table}
\label{table:A4dec}
\begin{tabular}{|c|c|}
\hline
$\mathcal{I}$ ($A_5$)& $\mathcal{A}_4$\\
\hline
$1$& $1$\\ 
$3$ &$3$\\
$3^{\prime}$ &3\\
$4$ &$3\oplus 1$\\
$5$& $3\oplus 1^\prime \oplus1^{\prime \prime }$\\
\hline
\end{tabular}
\caption{The decomposition of the irreducible representations of  the rotational icosahedral symmetry group $\mathcal{I}$ in terms of the irreducible representations of $\mathcal{A}_4$;  see also \cite{cumminspatera}.}
\end{table}

\noindent {\bf Group Presentation and the Shirai Basis.}  As is the case for any finite group, the elements of the rotational icosahedral symmetry group can be generated by a set of basic elements which satisfy certain relations; the elements and rules together are known as the ``presentation" of the group.  Unlike the case for smaller groups (such as $\mathcal{A}_4$), there are several equivalent ways to present the group, which either involve two generators or three generators; a classification can be found in \cite{Coxeter}.  We will proceed here by choosing a specific presentation of $\mathcal{I}$ due to Hamilton (see e.g.~\cite{Coxeter}):
\begin{equation}
\label{Hamilton}
<a,b| a^2=b^3=(ab)^5=e>.
\end{equation}
This notation indicates that with the two generators, \textit{a} and \textit{b},  the entire rotational icosahedral group can be generated (or something isomorphic to $\mathcal{I}$) when $a$ and $b$ are combined subject to the ``rules" of Eq.~(\ref{Hamilton}).  Here $e$ is the identity, as is standard in the mathematical literature.  An explicit example of this presentation is given in \cite{luhn}.

The two generators of Eq.~(\ref{Hamilton}) belong to the $C_2$ and $C_3$ conjugacy classes, as they square and cube to the identity, respectively.  We have found that for the purposes of flavor model building with the golden ratio prediction for the solar angle in mind, it is advantageous to use a presentation which explicitly uses an element either from $C_5$ or $C_5^2$, as the golden ratio appears in the character table for these conjugacy classes explicitly (see Table I).   As noted e.g.~in \cite{hoyle}, Hamilton's presentation can be rewritten in terms of an order two generator $S=a$ and an order five generator $T=bab$, such that  
\begin{equation}
<S,T|S^2=T^5=(T^2 S T^3 ST^{-1}STST^{-1})^3=e>.
\end{equation}
Explicit representations were presented both by Hoyle \cite{hoyle} and  Shirai \cite{shirai}. Hoyle's basis \cite{hoyle} is not unitary for a subset of the irreducible representations, and hence is not useful for flavor model building.  In Shirai's basis, which incorporates generators from $C_5$ and $C_2$ \cite{shirai}, the generators for the triplet representation ($3$) are as follows:
\begin{eqnarray}
\label{stmat3}
S_{3}=\frac{1}{2}\left( \begin{array}{ccc}
 -1 &\phi& \frac{1}{\phi}\\
 \phi&\frac{1}{\phi}& 1\\
 \frac{1}{\phi}&1&-\phi
 \end{array} \right ), \;\;\;\;
%\end{eqnarray}
%\begin{eqnarray}
T_{3}=\frac{1}{2}\left( \begin{array}{ccc}
 1&\phi&\frac{1}{\phi}\\
 -\phi&\frac{1}{\phi}& 1\\
\frac{1}{\phi}&-1&\phi
 \end{array} \right ),
\end{eqnarray}
while for the other independent triplet representation ($3^\prime$),  the generators take the form
\begin{eqnarray}
\label{stmat3p}
 S_{3^\prime}=\frac{1}{2}\left( 
\begin{array}{ccc}
 -\phi& \frac{1}{\phi}&1\\
 \frac{1}{\phi}&-1& \phi\\
 1&\phi&\frac{1}{\phi}
 \end{array} 
\right ),\;\;\;\;
T_{3^\prime}=\frac{1}{2}\left( \begin{array}{ccc}
 -\phi&-\frac{1}{\phi}&1\\
 \frac{1}{\phi}&1& \phi\\
- 1&\phi&-\frac{1}{\phi}
 \end{array} \right ).
\end{eqnarray}
The generators for the four-dimensional representation in the Shirai basis are given by
\begin{eqnarray}
\label{stmat4}
S_4=\frac{1}{4}\left(
\begin{array}{cccc}
 -1 & -1& -3& -\sqrt{5} \\
 -1 & 3& 1 & -\sqrt{5} \\
 -3 &1 & -1 & \sqrt{5} \\
 -\sqrt{5} & -\sqrt{5}& \sqrt{5}& -1
\end{array}
\right),\;\;\;
T_4=\frac{1}{4}\left(
\begin{array}{cccc}
 -1 & 1& -3 & \sqrt{5}\\
 -1 & -3 & 1 &\sqrt{5} \\
  3 & 1 &1 & \sqrt{5} \\
 \sqrt{5}& -\sqrt{5}& -\sqrt{5}& -1
\end{array}\right),
\end{eqnarray}
while for the five-dimensional irreducible representation, the order two generator is
\begin{eqnarray}
\label{smat5}
S_5=\frac{1}{2}\left(
\begin{array}{ccccc}
 \frac{1-3 \phi}{4}  & \frac{ \phi^2}{2} & -\frac{1}{2\phi^2}  & \frac{\sqrt{5}}{2} & \frac{\sqrt{3} }{4\phi} \\
 \frac{\phi^2}{2} & 1 & 1 & 0 & \frac{ \sqrt{3}}{2\phi}\\
 -\frac{1}{2\phi^2}  & 1& 0 & -1 & -\frac{\sqrt{3}\phi}{2} \\
 \frac{\sqrt{5}}{2} & 0 & -1 & 1 & -\frac{\sqrt{3}}{2} \\
 \frac{\sqrt{3}}{4\phi}   & \frac{\sqrt{3}}{2\phi}  & -\frac{ \sqrt{3}\phi}{2} &
-\frac{\sqrt{3}}{2} & \frac{3\phi-1}{4}
\end{array}\right),
\end{eqnarray}
and the order five generator is
\begin{eqnarray}
\label{tmat5}
T_5= \frac{1}{2}\left(
\begin{array}{ccccc}
  \frac{1-3 \phi}{4} &- \frac{\phi^2}{2}  & -\frac{1}{2\phi^2} & -\frac{\sqrt{5}}{2} & \frac{\sqrt{3}}{4\phi}
\\
 \frac{\phi^2}{2}  & -1 & 1 & 0 & \frac{\sqrt{3}}{2\phi} \\
 \frac{1}{2\phi^2}   & 1& 0 & -1 & \frac{\sqrt{3}\phi}{2}  \\
 -\frac{\sqrt{5}}{2} & 0 & 1 & 1 & \frac{\sqrt{3}}{2} \\
 \frac{ \sqrt{3}}{4\phi} & -\frac{\sqrt{3}}{2\phi} & -\frac{ \sqrt{3}\phi }{2} &
\frac{\sqrt{3}}{2} & \frac{3\phi-1}{4} 
\end{array}\right)
\end{eqnarray}
(our Eqs.~(\ref{smat5})--(\ref{tmat5}) correct minor typos found in \cite{shirai}).
Before proceeding further, let us first comment on other presentations that are available in the literature. The icosahedral ($\mathcal{A}_5$) group can be defined using three generators, in which two of the generators define the tetrahedral ($\mathcal{A}_4$) subgroup.  Such a presentation is given in the classic text of Lomont \cite{Lomont} and the work of Cummins and Patera \cite{cumminspatera}, as follows:
\begin{equation}
<A_1,A_2,A_3|A_1^3=A_2^2=A_3^2=(A_1A_2)^3=(A_2A_3)^3=(A_1A_3)^2=e>.
\end{equation}
In the above, the generators $A_1$ and $A_2$ generate the tetrahedral subgroup of $\mathcal{I}$.  The explicit representations, which we will not state explicitly here, differ in \cite{Lomont} and \cite{cumminspatera} by a similarity transformation.  In terms of the generators of Hamilton's presentation $a$ and $b$, we have the following equivalence relations (up to a similarity transformation):
\begin{eqnarray}
A_1\simeq  [b(ab^2)^2ab(ab^2)^2]^2,\;\;\; 
A_2\simeq  a, \;\;\;
A_3\simeq  bab^2a[b(ab^2)^2a]^2[b(ab^2)^2]^2.
\end{eqnarray}
While any of these presentations can be used as a basis for flavor model building, the Shirai basis has the advantage that the golden ratio appears explicitly, which will turn out to be a useful feature for seeking the golden ratio prediction for the solar mixing angle as given in Eq.~(\ref{goldenprediction}).  Therefore, in this paper we will use the Shirai basis (Eqs.~(\ref{stmat3})--(\ref{tmat5})).  The explicit tensor product decomposition has not, to our knowledge, been presented in the literature for this basis. We construct it explicitly in the next section. \\

\noindent {\bf  Tensor products.}
The next task is to find group invariants of $\mathcal{I}$.  General procedures for constructing the invariants for finite groups can be found in \cite{Hamermesh,Lomont}, with explicit applications to icosahedral ($\mathcal{A}_5$) symmetry in \cite{cumminspatera,luhn}.   In this paper, we will focus for the sake of simplicity on constructing the relevant Kronecker products within the Shirai basis.  Our results are sufficient for flavor model building at tree level using this basis.  Let  us begin by defining the two distinct triplet states as $3=(a_1,a_2,a_3)^T$ and $3=(b_1,b_2,b_3)^T$.  Using Eqs.~(\ref{stmat3})--(\ref{tmat5}), it is straightforward to show that the decomposition of $3 \otimes 3=1\oplus 3\oplus 5$ yields the following results for the singlet,
\begin{equation}
\label{singlet33}
1= a_1 b_1 + a_2 b_2 + a_3 b_3,
\end{equation}
the triplet,
\begin{eqnarray} 
3= (a_3 b_2 - b_2 a_3,a_1 b_3- a_3 b_1,a_2 b_1 -a_1 b_2)^T,
\end{eqnarray}
and the fiveplet:
\begin{eqnarray}\nonumber
5= 
\left(
\begin{array}{c}
a_2 b_2-a_1 b_1\\
a_2 b_1 +a_1 b_2\\
a_3 b_2 + a_2 b_3 \\
a_1 b_3 + a_3 b_1\\
-\frac {1}{\sqrt{3}}( a_1 b_1 +a_2 b_2 -2a_3 b_3)
\end{array} \right),
\end{eqnarray}
which show the symmetry of the singlet and the fiveplet, and the antisymmetry of the triplet.  The Kronecker products for $3^\prime\otimes 3^\prime$ for the singlet and the triplet have an identical structure to those of the $3\otimes 3$, but for the fiveplet we have the different symmetric combination:
\begin{eqnarray}\nonumber
5= 
\left(
\begin{array}{c}
\frac{1}{2}(-\frac{1}{\phi}a_1 b_1-\phi a_2b_2+\sqrt{5}a_3b_3)\\
a_2 b_1 +a_1 b_2\\
-(a_3 b_1 + a_1 b_3) \\
a_2 b_3 + a_3 b_2\\
\frac {1}{2\sqrt{3}}( (1-3\phi)a_1 b_1 +(3\phi-2)a_2 b_2 +a_3 b_3)
\end{array} \right).
\end{eqnarray}
For $3 \otimes 3^{\prime}=4 \oplus 5$, we have the following result for the fourplet:
\begin{eqnarray}
4= \left(
\begin{array}{c}
\frac{1}{\phi}a_3 b_2-\phi a_1 b_3\\
\phi a_3 b_1+ \frac{1}{\phi}a_2 b_3\\
-\frac{1}{\phi}a_1 b_1+\phi a_2 b_2\\
a_2 b_1 -a_1 b_2 + a_3 b_3
\end{array} \right),
\end{eqnarray}
and the five-dimensional representation takes the form:
\begin{eqnarray}
5=\left(
\begin{array}{c}
\frac{1}{2}(\phi^2a_2 b_1+\frac{1}{\phi^2}a_1b_2-\sqrt{5}a_3b_3)\\
-(\phi a_1 b_1+\frac{1}{\phi} a_2 b_2)\\
\frac{1}{\phi}a_3 b_1-\phi a_2 b_3\\
\phi a_3 b_2+ \frac{1}{\phi}a_1 b_3\\
\frac{\sqrt{3}}{2}(\frac{1}{\phi} a_2 b_1+\phi a_1 b_2 +a_3 b_3)
\end{array} \right).
\end{eqnarray}
For our purposes in this paper, the only remaining Kronecker products we need are (i) the singlet which results from tensoring together two fourplets, $4=(a_1,a_2,a_3,a_4)^T$ and $4=(b_1,b_2,b_3,b_4)^T$, which is given by
\begin{equation}
1= a_1 b_1 + a_2 b_2 + a_3 b_3+a_4b_4,
\end{equation}
and (ii) the singlet which results from the tensor decomposition of the direct product of two fiveplets, $5=(a_1,a_2,a_3,a_4,a_5)^T$ and $5=(b_1,b_2,b_3,b_4,b_5)^T$, which takes the familiar form
\begin{equation}
1 = a_1 b_1 + a_2 b_2 + a_3 b_3+a_4b_4+a_5b_5.
\end{equation}
The form of these singlets results directly from unitarity and the reality of the Shirai basis.  In forthcoming work \cite{alexlisalong}, we will present a more comprehensive analysis of the group invariants which connects with the general formalism presented in \cite{cumminspatera,luhn}.

\section{Lepton Flavor Model Building}
\label{sIII}
In this section, we apply this formalism to the issue of flavor model building based on the rotational icosahedral symmetry group as a family symmetry of the lepton sector. We focus in this paper exclusively on the lepton sector and defer the quark sector to future work.   With this restriction in mind, the starting point is to examine possible family symmetry charge assignments for the SM fields, which include the lepton doublets $L_i$, the charged lepton singlets $\bar{e}_i$, and, in the case of seesaw models, the right-handed neutrinos $\bar{N}_i$ ($i$ is a family index).   The electroweak Higgs field(s) are assumed to be blind to the family symmetry. Clearly, it is natural for the lepton sector fields to transform as a $3$ or a $3^\prime$ under the icosahedral symmetry.  This leads to eight possible scenarios, as shown in Table IV.
\begin{table}
\label{table:charges}
\begin{tabular}{|c|cccccccc|}
\hline
$L$& 3&3&3&$3^\prime$&3&$3^\prime$&$3^\prime$&$3^\prime$\\
\hline
$\bar{e}$& 3&3&$3^\prime$&3&$3^\prime$&3&$3^\prime$&$3^\prime$\\
\hline
$\bar{N}$& 3&$3^\prime$&3&3&$3^\prime$&$3^\prime$&3&$3^\prime$\\
\hline
\end{tabular}
\caption{Possible charge assignments under $\mathcal{I}$ for the SM lepton doublets $L$, lepton singlets $\bar{e}$, and the right-handed neutrinos $\bar{N}$.}
\end{table}

For concreteness, we will focus here on the case in which $L$ is a $3$ and $\bar{e}$ is a $3^\prime$, leaving the $\bar{N}$ charge unspecifed for the moment.  With these charge assignments, the standard procedure is to use the Froggatt-Nielsen mechanism \cite{fn} and introduce flavon fields, scalar fields charged under the family symmetry which acquire vacuum expectation values (vev's) that break the family symmetry.   To determine the options for the flavon field content, recall that the structure of the mass terms we aim to generate takes the standard schematic form:
\begin{equation}
\label{massL}
-\mathcal{L}_m=Y^{(\nu)}_{ij} L_i \bar{N}_j  H+ M^{(M)}_{ij}\bar{N}_i \bar{N}_j +Y^{(e)}_{ij} L_i \bar{e}_j  H +{\rm h.c.},
\end{equation}
in which $Y_{ij}$ denote effective Yukawa couplings, $M^{(M)}_{ij}$ represents the Majorana mass matrix of the right-handed neutrinos, and $H$ denotes the electroweak Higgs field.   (For simplicity, we assume a single Higgs field, as in the SM; it is straightforward to generalize the analysis for the case of two  electroweak Higgs doublets, as in the minimal supersymmetric extension of the SM.)  The Dirac mass matrices $M^{(f)}_D$ are then given by $Y^{(f)} \langle H \rangle$.   Within the seesaw framework, the eigenvalues of $M^{(M)}_{ij}$ are much larger than the Higgs vacuum expectation value, such that the heavy right-handed neutrinos can be integrated out to yield the standard seesaw mass matrix for the light (primarily left-handed) neutrinos \cite{seesaw}:
\begin{equation}
M_\nu=M_D (M^{(M)})^{-1} M_D^T.
\end{equation}
This mass term can also be encoded by Weinberg's dimension five operator \cite{weinberg}:
\begin{equation}
\label{dim5}
-\mathcal{L}_w= \frac{a_{ij}}{M} L_i H L_j H,
\end{equation}
where $M$ is a (presumably high) cutoff scale ({\it i.e.}, the scale of the eigenvalues of the Majorana mass matrix $M^{(M)}$ of the right-handed neutrinos in a seesaw framework). For simplicity, we will use Eq.~(\ref{dim5}) to encode the effective seesaw mass term, which bypasses the details of the right-handed neutrinos, for the time being.  We will return to this issue later in the paper.   

In our scenario, the effective mass terms are thus of the $LL$ and $L\bar{e}$ type, with 
\begin{equation}
LL: 3\otimes 3=1\oplus 3 \oplus 5,\;\;\;\;\; L\bar{e}: 3\otimes 3^\prime = 4\oplus 5.
\end{equation}
We see that in the absence of flavor symmetry breaking, the bare $LL$ term is allowed, but the bare $L\bar{e}$ term is absent.  Given the structure of the singlet coupling of two triplets (see Eq.~(\ref{singlet33})), this indicates that the neutrinos can have degenerate masses of order $\langle H \rangle^2/M$ and the charged leptons are all massless at leading order.  This result is, of course, quite far from the observed pattern of lepton flavor mixing.  However, the flavor structure emerges through the spontaneous breaking of the icosahedral symmetry through the vacuum expectation values of flavon sector fields. This is a generic feature for flavor models based on non-Abelian discrete family symmetries.   The details of the flavon sector matter content and couplings are thus the key elements to consider for flavor model building.  Within icosahedral symmetry, we see that the flavon sector fields of interest are singlets and fiveplets for the neutrino sector, and fourplets and fiveplets in the charged lepton sector (the symmetry of the $LL$ term disallows the coupling of $LL$ to a flavon field which transforms as a triplet).

Even within this simpler class of scenarios in which the $L$ and $\bar{e}$ fields have been assigned to specific representations, to address whether the golden ratio prediction for the solar mixing angle can naturally emerge within such icosahedral ($A_5$) models requires a systematic study, which is currently in progress \cite{alexlisalong}.  Here, for illustrative purposes, we will focus on constructing a minimal (toy) scenario which reproduces a similar structure of mixings as found in  \cite{Kajiyama:2007gx}, in which the solar mixing angle originates from the neutrino sector, and the maximal atmospheric mixing angle arises from the charged lepton sector.  In this paper, we do not attempt to construct a fully realistic model; instead, the goal is to see, through this very simple example, what flavon sector fields and associated vacuum expectation values are needed to reproduce this pattern within the framework of icosahedral symmetry.  To this end, we include a ``minimal" number of flavon fields and only consider leading order couplings.  Furthermore,  we will not include effects of renormalization group running in this scenario, which would be necessary in a comprehensive study (see e.g.~\cite{Antusch:2003kp}).

The simple model we consider has the following flavon sector fields: the fields $\xi$ and $\psi$, which each transform as a $5$, and the flavon $\chi$, which transforms as a $4$.  The presence of two distinct fiveplets turns out to be unavoidable in this context; we will comment on this shortly.  We will assume that  $\xi$ couples to the $LL$ operator and that  $\chi$ and $\psi$ couple to the $L\bar{e}$ operator; this assumption necessitates additional symmetries which forbid $\xi$ and $\psi$ from coupling to $L\bar{e}$ and  $LL$, respectively.   For simplicity, we will also assume that the tree level $LL$ term is forbidden by such additional symmetries.  
This tree level term is proportional to the identity matrix, and hence it does not affect the mixing angles or the predictions for the mass-squared differences, but does affect the overall neutrino mass scale.  Similarly, singlet flavon fields which couple to $LL$ would also just affect the overall neutrino mass scale.  We assume such singlet fields are absent.

With these assumptions, the mass terms take the following schematic form:
\begin{equation}
\label{lmass}
-\mathcal{L}_{mass}= \frac{\alpha_{ijk}}{M M^\prime}L_iHL_jH\xi_k+\frac{\beta_{ijk}}{M^\prime}L_i\bar{e}_jH\psi_k+\frac{\gamma_{ijl}}{M^\prime}L_i\bar{e}_jH\chi_l +{\rm h.c.},
\end{equation}
in which $M^\prime$ represents the (presumably high) scale of flavor symmetry breaking, and $\alpha_{ijk}$, $\beta_{ijk}$, and $\gamma_{ijl}$ are dimensionless couplings that encode the tensor product decomposition of the icosahedral symmetry as well as any (presumably $O(1)$) coupling factors.    In principle, $M^\prime$ has no relation to $M$, particularly within the context of the neutrino seesaw.  After electroweak symmetry breaking, the resulting mass matrices are given by
\begin{equation}
\label{neutrinotemplate}
M_\nu= \alpha \left(
\begin{array}{ccc}
-\xi_1-\frac{1}{\sqrt{3}}\xi_5 & \xi_2 &\xi_4\\
\xi_2& \xi_1-\frac{1}{\sqrt{3}}\xi_5 & \xi_3\\
\xi_4 & \xi_3& \frac{2}{\sqrt{3}} \xi_5
\end{array}\right),\;\;\;
\end{equation}
\begin{eqnarray}
\label{chgleptemplate}
M_e&=& \beta \left ( \begin{array}{ccc} - \phi \psi_2& \frac{1}{2\phi^2} \psi_1+\frac{\sqrt{3}\phi}{2} \psi_5 & \frac{1}{\phi}\psi_4\\
\frac{\phi^2}{2}\psi_1+\frac{\sqrt{3}}{2\phi}\psi_5 &-\frac{1}{\phi}\psi_2&-\phi \psi_3\\
\frac{1}{\phi}\psi_3 & \phi \psi_4 &-\frac{\sqrt{5}}{2}\psi_1+\frac{\sqrt{3}}{2}\psi_5  \end{array} \right )  
+\gamma \left ( \begin{array}{ccc}
-\frac{1}{\phi} \chi_3 &-\chi_4&-\phi \chi_1\\
\chi_4&\phi \chi_3&\frac{1}{\phi}\chi_2\\
\phi \chi_2&\frac{1}{\phi} \chi_1&\chi_4
\end{array}\right),\nonumber \\
\end{eqnarray}
in which $\alpha$, $\beta$, and $\gamma$ are dimensionless couplings (functions of the parameters of Eq.~(\ref{lmass})) which encode the suppression of the neutrino and the charged lepton mass scales.

To determine the nature of the flavor symmetry breaking and the resulting masses and mixings,  the dynamics of the flavon sector must be considered in detail.  The flavon structure is clearly quite rich due to the presence of the four- and five-dimensional representations.  Given the complexity of this sector, we will leave the flavon dynamics unspecified in this paper as a first step toward flavor model building, and defer this important issue for future work.  We will instead examine the possible vev patterns that can result in scenarios analogous to those presented in \cite{Kajiyama:2007gx}, and defer a more systematic study of other viable options to future work.  Within this approach,  the golden prediction for the solar neutrino mixing angle can be obtained in the neutrino sector by assigning the following vev to $\xi$:
\begin{eqnarray}
\label{xivev}
\left(
\begin{array}{c}
\xi_1\\
\xi_2\\
\xi_3\\
\xi_4\\
\xi_5\\
\end{array} \right)
\longrightarrow 
\frac{\sqrt{3}}{2\alpha}\left(
\begin{array}{c}
\frac{1}{\sqrt{15}}(m_2-m_1)\\
\frac{2}{\sqrt{15}}(m_2-m_1)\\
0\\
0\\
-(m_1+m_2)
\end{array}\right).
\end{eqnarray}
In the above, $m_1$ and $m_2$ are mass parameters, which turn out to be the neutrino mass eigenvalues associated with the solar neutrino pair.   The dominant component of the vacuum expectation value of $\xi$ is that of $\xi_5$; the other terms represent small corrections, of order $\sim \Delta m^2_{12}/(m_1+m_2)$.  From Eq.~(\ref{neutrinotemplate}), one sees that the vanishing of the $\xi_{3,4}$ component vevs is needed to ensure that only the solar mixing angle results from the neutrino sector.  The solar angle in this scenario is determined by the interplay between the $\xi_{1}$ and $\xi_2$ vevs, as the $\xi_5$ vev leads to a diagonal mass matrix.
 %ADD MORE HERE.
Explicitly, the neutrino mass matrix is
\begin{eqnarray}
\label{mnumat}
M_\nu =\frac{1}{\sqrt{5}}\left(
\begin{array}{ccc}
\phi m_1+\frac{1}{\phi} m_2&m_2-m_1 &0\\
m_2-m_1& \frac{1}{\phi} m_1+\phi m_2&0\\
0&0&-\sqrt{5}(m_1+m_2)
\end{array}
\right).
\end{eqnarray}
The neutrino mass eigenvalues are $m_1$, $m_2$, and $m_3=-(m_1+m_2)$.  The constraint on $m_3$ results from the interplay of the small $\xi_1$ and $\xi_2$ vevs with the $\xi_5$ vev, which is needed to obtain the golden ratio prediction for the solar mixing angle. The negative sign of $m_3$ can be absorbed by a phase rotation because the neutrinos are Majorana fermions. The neutrino mixing matrix $U_\nu$, defined by $U_\nu M_\nu U_\nu^T$, satisfies Eq.~(\ref{goldenprediction}) by design:
\begin{eqnarray}
\label{unumat}
U_{\nu}=\left(
\begin{array}{ccc}
\sqrt{\frac{\phi}{\sqrt{5}}} & -\sqrt{\frac{1}{\sqrt{5}\phi}} & 0\\ \sqrt{\frac{1}{\sqrt{5}\phi}} & \sqrt{\frac{\phi}{\sqrt{5}} }& 0\\
0&0&-i \end{array}\right ).
\end{eqnarray}
The masses $m_1$ and $m_2$ are determined by accommodating the splittings of the solar and atmospheric neutrino mass-squares (see Eqs.~(\ref{msqsol})--(\ref{msqatm})); we obtain $m_1=2.77 \times 10^{-2}$ eV, $m_2=2.91\times 10^{-2}$ eV, and $\vert m_3 \vert =5.68 \times 10^{-2}$ eV. Hence, the resulting pattern has a normal mass ordering, but with a mass spectrum which is quasi-degenerate.%~\footnote{Running effects may then be significant; as stated, they are neglected here in our analysis of this toy scenario for simplicity.}   
The mass pattern obtained here differs from that of \cite{Kajiyama:2007gx}, in which $m_3$ was an independent parameter, while $m_1$ and $m_2$ were given by $m_1=m/\phi$, $m_2=\phi m$ ($\Delta m^2_{12}=\sqrt{5}m^2$ in their scenario).

In the charged lepton sector, we seek maximal mixing between the second and third families.  At leading order in flavon fields, we assume only $m_\tau$ is nonvanishing, and 
that the charged lepton mass matrix takes the form 
\begin{equation}
\label{strumiachgleplead}
M_{e}=\frac{1}{\sqrt{2}} \left ( \begin{array}{ccc} 0 &0&0 \\ 0& 0 &m_\tau \\ 0&0 & m_\tau \end{array} \right ),
\end{equation}
as in \cite{Kajiyama:2007gx}.  Such a pattern can be achieved with the following vev assignments for $\chi$ and $\psi$:  
\begin{eqnarray}
\label{psivev}
\left(
\begin{array}{c}
\psi_1\\
\psi_2\\
\psi_3\\
\psi_4\\
\psi_5\\
\end{array}\right)
\longrightarrow
\frac{m_\tau}{2\sqrt{6}\beta }\left(
\begin{array}{c}
- \sqrt{\frac{5}{3}}\\
 0\\
-\frac{2}{\sqrt{3}} \phi \\
  0\\
 1 \\ 
\end{array} \right),
\end{eqnarray}
\begin{eqnarray}
\label{chivev}
\left(
\begin{array}{c}
\chi_1\\
\chi_2\\
\chi_3\\
\chi_4\\
\end{array} \right)
\longrightarrow
\frac {m_\tau}{3\sqrt{2}\gamma}\left(
\begin{array}{c}
0\\
 \frac{1}{\phi}\\
 0\\
 1
\end{array}\right).
\end{eqnarray}
Given Eq.~(\ref{strumiachgleplead}), the vevs of $\psi_{2,4}$ and $\chi_{1,3}$ are required to vanish due to the structure of Eq.~(\ref{chgleptemplate}). The left-handed states are diagonalized by the mixing matrix
\begin{eqnarray}
\label{uemat}
U_{e}=\left(
\begin{array}{ccc}
1&0&0\\
0&\frac{1}{\sqrt{2}}&-\frac{1}{\sqrt{2}}\\
0&\frac{1}{\sqrt{2}}&\frac{1}{\sqrt{2}}\\
\end{array}\right).
\end{eqnarray}
It can be shown that Eq.~(\ref{strumiachgleplead})  cannot be reproduced with nonvanishing vevs for only $\chi$ or only $\psi$ (see Eq.~(\ref{chgleptemplate})).   Furthermore, two distinct fiveplets are required in this model (one which couples to the neutrino sector and one which couples to the charged lepton sector), since the required vev of $\psi$ is qualitatively different than that of $\xi$ (see Eq.~(\ref{xivev})). 
% Both of these features are drawbacks of this simple model.

Nonvanishing electron and muon masses can also be accommodated by fitting the full result for the charged lepton mass matrix given in \cite{Kajiyama:2007gx}:
\begin{equation}
\label{strumiachglep}
M_{e}=\frac{1}{\sqrt{2}} \left ( \begin{array}{ccc} \sqrt{2} m_e &0&0 \\ 0& m_\mu &m_\tau \\ 0&-m_\mu & m_\tau \end{array} \right ),
\end{equation}
by allowing for nonvanishing vevs for $\psi_{2,4}$ and $\chi_{1,3}$ as follows:
\begin{eqnarray}
\left(
\begin{array}{c}
\psi_1\\
\psi_2\\
\psi_3\\
\psi_4\\
\psi_5\\
\end{array}\right)
\longrightarrow
\frac{1}{2\sqrt{6}\beta }\left(
\begin{array}{c}
- \sqrt{\frac{5}{3}} m_\tau\\
 \frac{2}{\sqrt{3}} (-\phi \sqrt{2} m_e-\frac{1}{\phi} m_\mu)\\
-\frac{2}{\sqrt{3}} \phi m_\tau\\
 -\frac{2}{\sqrt{3}} \phi m_\mu \\
 m_\tau \\ 
\end{array} \right),
\end{eqnarray}
\begin{eqnarray}
\left(
\begin{array}{c}
\chi_1\\
\chi_2\\
\chi_3\\
\chi_4\\
\end{array} \right)
\longrightarrow
\frac {1}{3\sqrt{2}\gamma}\left(
\begin{array}{c}
-\frac{1}{\phi}m_\mu  \\
\frac{1}{\phi}m_\tau\\
-\frac{\sqrt{2}}{\phi}m_e+\phi m_\mu \\
 m_\tau
\end{array}\right).
\end{eqnarray}
Such small entries in the flavon vevs clearly represent a large fine-tuning, which is needed to achieve the hierarchically suppressed electron and muon masses at leading order in flavon fields. In general, whether or not flavon vevs such the ones presented above can emerge naturally within icosahedral symmetry is a question which can only be addressed by explicitly considering the dynamics of the flavon sector.  Though the vacuum alignment issue must be addressed, it is not implausible that the appearance of the golden ratio in the flavon vevs can emerge naturally from the underlying icosahedral symmetry.

We pause here to note that generating charged lepton masses and mixing is quite different here than in the case of $\mathcal{A}_4$ models, where typically the charged lepton singlets are assigned to the three different one-dimensional irreducible representations.   In such $\mathcal{A}_4$ models, the hierarchical charged lepton masses can be addressed via separate small Yukawa couplings, whereas in the $\mathcal{I}$ ($\mathcal{A}_5$) scenarios studied here,  one necessarily has to address all three families together.  In forthcoming work, we will analyze details of higher-order couplings which can result in the charged lepton mass matrix of Eq.~(\ref{strumiachglep}), among other possibilities.

The last feature remaining is to present the explicit form of the MNSP matrix.   Given Eq.~(\ref{unumat}) and Eq.~(\ref{uemat}), the MNSP matrix takes the form
\begin{equation}
\label{MNSP}
U=U_{\rm MNSP}=U_e U_\nu^\dagger =\left ( \begin{array}{ccc} 
\sqrt{\frac{\phi}{\sqrt{5}}} & \sqrt{\frac{1}{\sqrt{5}\phi}} & 0\\ -\frac{1}{\sqrt{2}}\sqrt{\frac{1}{\sqrt{5}\phi}} & \frac{1}{\sqrt{2}}\sqrt{\frac{\phi}{\sqrt{5}} }& -\frac{1}{\sqrt{2}}\\
-\frac{1}{\sqrt{2}}\sqrt{\frac{1}{\sqrt{5}\phi}} & \frac{1}{\sqrt{2}}\sqrt{\frac{\phi}{\sqrt{5}} }& \frac{1}{\sqrt{2}}\\
\end{array} \right ) \mathcal{P},
\end{equation}
where $\mathcal{P}={\rm Diag}(1,1,i)$ is the Majorana phase matrix.
By design, the MNSP matrix has a vanishing reactor mixing angle, a maximal atmospheric mixing angle, and a solar angle given by $\theta_{12}={\rm tan}^{-1}(1/\phi)$, as in Eq.~(\ref{goldenprediction}).

In this scenario, once the charges of $L$ and $\bar{e}$ are specified and the flavon field content is chosen, there are in principle fourteen complex parameters (the flavon vevs) which are inputs for the model.  As there are eight physical observables (3 charged lepton masses, two neutrino mass-squared differences, and three lepton mixing angles), without further input the model is underconstrained.  With the choice of flavon field vevs of Eqs.~(\ref{xivev}), (\ref{psivev}), and (\ref{chivev}), this number of parameters has been reduced by hand to 8 real parameters. Therefore, we can predict observables such as the neutrinoless double beta decay parameter, $m_{\beta\beta}=\sum_i m_i U_{ei}^2$, which is given by
\begin{equation}
m_{\beta\beta}=\frac{m_1\phi}{\sqrt{5}}+\frac{m_2}{\phi\sqrt{5}}.
\end{equation}
In our model, $m_{\beta\beta}=2.8\times 10^{-2}$ eV. In regards to CP phases, the Dirac MNSP phase cannot be predicted because $\theta_{13}=0$ to this order of approximation.  As seen in  Eq.~(\ref{MNSP}), our scenario does have a Majorana phase, but it does not enter into the expression for $m_{\beta\beta}$.

We close this section by commenting on seesaw realizations of scenarios such as the one presented above.  Given the form of Eq.~(\ref{massL}), we see that whether the right-handed neutrinos are assigned to transform as a $3$ or a $3^\prime$, the possible flavon fields in the right-handed neutrino sector are either a $1$ or a $5$ due to the symmetry of the $\bar{N}\bar{N}$ operator.  One straightforward example for the case in which $\bar{N}$ is a $3$ (more precisely, when $L$ and $\bar{N}$ transform identically under $\mathcal{I}$) is that the only flavon field which couples to $L\bar{N}$ and  $\bar{N}\bar{N}$ is the fiveplet field $\xi$.  We will also assume that the tree-level coupling of $\bar{N}\bar{N}$ is forbidden (e.g. by an additional symmetry).  In this case, $M^{(M)}$ is proportional to the Dirac mass matrix, $M_D$, which results from the $L\bar{N}$ coupling.  The seesaw formula then results in 
\begin{equation}
M_\nu=M_D (M^{(M)})^{-1} M_D^T\simeq M_D M_D^{-1} M_D^T \simeq M_D,
\end{equation}
since $M_D$ is a symmetric matrix in this case.  If $\xi$ has the same vev pattern as in Eq.~(\ref{xivev}), it is straightforward to see that we obtain exactly the same flavor mixing structure of the neutrino mass matrix of Eq.~({\ref{mnumat}).  Therefore, this scenario represents a simple seesaw extension of the effective theories given above.  However, it is clear that there are many other possible seesaw embeddings that are worthy of further exploration.   We plan to analyze this and other crucial model-building issues, most notably the dynamics of the flavon sector, in future work.

\section{Conclusions}
In this paper, we have investigated the possibility that the rotational icosahedral symmetry group $\mathcal{I}$, which is isomorphic to $\mathcal{A}_5$, is the family symmetry group of the lepton sector.  Icosahedral symmetry has remained relatively unexplored for physics applications, most notably because it is not compatible with translational invariance and thus is not a crystallographic point group. Nevertheless, the rotational icosahedral symmetry group has many features which make it an appealing candidate for a family symmetry.  In particular, the presence of two distinct triplet irreducible representations provides a rich framework in which to generate the large mixing angles of the lepton sector.

We have focused in this paper on developing the tools to construct models in which the solar mixing angle is related in a nontrivial way to the golden ratio, $\phi=(1+\sqrt{5})/2$.
Icosahedral symmetry provides a natural setting in which to explore this hypothesis, as the golden ratio emerges from the geometrical properties of the icosahedron.  
Although the golden ratio prediction for the solar mixing angle is an intriguing option, it is not the only plausible outcome in models with icosahedral symmetry.  One possible research direction is to study how tri-bimaximal mixing might arise within this framework.  This outcome is also in some sense natural, in that the tetrahedral ($\mathcal{A}_4$) group is a subgroup of $\mathcal{I}$.

Our work represents the first stage for flavor model building using icosahedral symmetry.  We have constructed explicit tensor products using a specific group representation due to Shirai \cite{shirai}, and used these results to construct toy models which reproduce the golden ratio prediction for the solar mixing angle of Eq.~(\ref{goldenprediction}).  Much work still needs to be done within this framework, including a more general analysis of higher-order invariants and the implications for flavor model building beyond leading order, particularly with respect to the important issue of the flavon sector dynamics.  
Though our investigation is still at a preliminary stage, icosahedral symmetry clearly provides a rich arena in which to explore the flavor puzzle of the Standard Model and is certainly worthy of further study in this exciting era of continued experimental probes of lepton flavor mixing.

\section{Acknowledgments}
We thank P.~Ramond, I.~W.~Kim, and D.~Chung for helpful discussions. This work was supported by the U. S. Department of Energy contract DE-FG-02-95ER40896.    

%\bibliographystyle{unsrt}
%\begin{references}
\bibliographystyle{prsty}

\end{document}